\documentclass[a4paper,12pt]{article}

\usepackage[latin2]{inputenc}
\usepackage{graphicx}
\usepackage{amstext}
\usepackage[]{ntheorem}
\usepackage{fancyhdr}
\usepackage{amsfonts}
\usepackage{amssymb}
\usepackage{amsmath,chemarrow}
\usepackage{array}
\usepackage{multirow}

\usepackage{t1enc}
\usepackage{tikz}
\usepackage{hyperref}
\usepackage[affil-it]{authblk}
\usepackage{pgfplots}
\usepackage{caption}
\usepackage{subcaption}
\usepackage{float}

\usepackage{makecell}

\newcommand{\Real}{\mathbb{R}}

\providecommand{\keywords}[1]
{
  \small	
  \textbf{\textit{Keywords---}} #1
}

\title{Bid-aggregation based clearing of day-ahead electricity markets}

\author[1]{Botond Feczk\'{o}
\protect
}
\author[2]{D\'{a}niel Div\'{e}nyi
\protect
}
\author[2]{\'{A}d\'{a}m Sleisz
\protect
}
\author[1,3]{D\'{a}vid Csercsik
\protect \footnotemark
}

\affil[1]{P\'{a}zm\'{a}ny P\'{e}ter Catholic University\\ Faculty of Information Technology and Bionics\\ Pr\'{a}ter~u. 50/A 1083 Budapest, Hungary \\
              Tel.: +36-1 886 47 00
              Fax: +36-1 886 47 24}

\affil[2]{Budapest University of Technology and Economics, Department of Electric Power Engineering\\ Egry J. u. 18. 1111 Budapest, Hungary \\
              Tel.: +36-1 463-2904
              Fax: +36 1 463-3600}  
              
\affil[3]{Centre for Economic and Regional Studies, Institute of Economics\\ T\'{o}th K\'{a}lm\'{a}n u. 4 1097 Budapest, Hungary \\
              Tel.: +36-1 224 6700 \\
              $^*$ Corresponding author, email: csercsik.david@krtk.hun-ren.hu}

\date{October 30, 2022}

\begin{document}

\maketitle

\abstract{In this work we propose a heuristic clearing method of day-ahead electricity markets. In the first part of the process, a computationally less demanding problem is solved using an approximation of the cumulative demand and supply curves, which are derived via the aggregation of simple bids. Based on the outcome of this problem, estimated ranges for the clearing prices of individual periods are determined.
In the final step, the clearing for the original bid set is solved, taking into account the price ranges determined previously as constraints. Adding such constraints reduces the feasibility region of the clearing problem. By removing simple bids whose acceptance or rejection is already determined by the assumed price range constraints, the size of the problem is also significantly reduced.
Via simple examples, we show that due to the possible paradox rejection of block bids the proposed bid-aggregation based approach may result in a suboptimal solution or in an infeasible problem, but we also point out that these pitfalls of the algorithm may be avoided by using different aggregation patterns.
We propose to construct multiple different aggregation patterns and to use parallel computing to enhance the performance of the algorithm. 
We test the proposed approach on setups of various problem sizes, and conclude that in the case of parallel computing with 4 threads a high success rate and a significant gain in computational speed may be achieved.
}


\newpage

\section{Introduction}
\label{sec1}
Day-ahead electricity markets (DAMs) are two-sided multi-unit auctions where market participants submit orders to buy or sell electric power in a given area during predefined periods of the following day. In the current paper we are addressing the problem of market clearing in the European-type framework of DAM, where self-scheduling units are assumed. In this setup, market players agree on a collection of rules describing the bid acceptance and the price determination. The market operator collects the participants' orders, and determines the dispatch in order to maximize the total social welfare (TSW), the total utility of consumption minus the total cost of production. In the presence of a general market clearing price, the total social welfare of any period may be decomposed to the welfare or surplus of single bids.

The computed market prices support a market equilibrium, which maximizes the social welfare \cite{madani2017revisiting}, with respect to the actual constraints (e.g. keeping production and consumption in balance for every period). 
The most difficult feature to deal with in day-ahead electricity markets is the fact, that some order types are non-convex \cite{madani2016non}.

Non-convexity originates from technical constraints such as minimum performance  \cite{sleisz2019new}, and from the cost structure (e.g. start-up costs) of generating units. 
In this study we restrain the analysis to the presence of \emph{block orders} \cite{meeus2009block}, which are the most common form of non-convex bids. Block orders may incorporate multiple periods and exhibit the \emph{fill-or-kill} property (they may be either entirely accepted or entirely rejected in all relevant periods). In the presence of such bids, it is necessary to allow paradox rejection (or paradox acceptance\footnote{The generally used approach is to allow paradox rejection to ensure the profitability of all accepted orders.}), to ensure the universal existence of market clearing prices (MCPs) \cite{madani2014minimizing}.

\subsection{Motivation}
\label{subsec_motivation}

The reference solution for market clearing of European-type DAMs \cite{chatzigiannis2016european_model} is the EUPHEMIA formalism and market coupling tool \cite{euphemia2015}, which operates since 2014 in the integrated power exchanges of Europe, and has been widely analyzed and researched in multiple aspects, including potential conceptual development \cite{chatzigiannis2016european,koltsaklis2018incorporating,koltsaklis2020assessing} and policy implications \cite{koltsaklis2018policy} since its introduction.

Although EUPHEMIA operates successfully for several years, the size and complexity of the practical clearing problems pose considerable challenges and require continuous research and development. The former aspect, the size of the problem naturally grows with the progress of the market-integration process. Furthermore, the pan-European DAM is expected to move from the time resolution of 60 minutes to 15 minutes, and this change might not be feasible with the existing formulation \cite{sdac2022}. In this light, any novel concept which may potentially improve the computational efficiency can be regarded as valuable. 

\subsection{Related literature}
\label{subsec_rel_lit}

The literature related to the topic of European-type DAMs is very extensive. For the review of the most important concepts, trends and open problems, one may refer to the papers  \cite{chatzigiannis2016european_model,tanrisever2020european,koltsaklis2020assessing}. In the current work we are restraining ourselves on the literature review of the narrower field, namely on approaches explicitly focusing on computationally efficient clearing solutions of DAMs.

The first computationally efficient algorithm designed for European DAMs was the COSMOS algorithm \cite{cosmos2011}. COSMOS constructs and solves a Mixed Integer Problem (MIP) where integer variables are introduced due to the fill-or-kill property of block orders. EUPHEMIA can be viewed as an extended version of COSMOS with additional functionality (e.g. several new order types). Madani et al. present an alternative MIP formulation based on duality theory and a solution algorithm using Benders-decomposition in \cite{madani2015computationally}, which proves to be computationally highly efficient.
The paper \cite{dourbois2018novel}, incorporating complex and 'Prezzo Unico Nazionale' (PUN) orders and proposing a Mixed complementarity problem formulation also aims to provide a computationally efficient formulation.
The paper \cite{ceyhan2022extensions} proposes an improved version of the Benders-decomposition algorithm to enhance efficiency, using the Turkish DAM as basis for testing.
Regarding the more unconventional approaches, recent works address also the Turkish DAM and propose a genetic algorithm-based approach \cite{sensoy2018genetic} and an adaptive tabu search algorithm \cite{kurt2018adaptive} for the market clearing.

\subsection{Contribution and outline}
\label{subsec_contr_outline}

In this work we propose a clearing concept, which is based on the subsequent solution of two clearing problems. Since the two clearing steps may be performed by any other innovative method (e.g. also in EUPHEMIA), the proposed approach may be combined with practically any of the aforementioned computationally efficient algorithms.
The basic idea of the proposed approach is based on the well-known geometric interpretation of the market clearing problem. If one considers only the most widely used bid types of day-ahead power exchanges (DAPXs), the one-period bids in the clearing, for which partial acceptance is allowed (so called \emph{standard bids}), the MCPs and thus the result of the clearing may be obtained by considering the intersection point of the cumulative supply and demand curves period-wise, which are derived by the sorting of bids by price (ascending in the case of supply and descending in the case of demand). Naturally, if non-convex orders, like block orders are also present, these cumulative curves of standard bids must be combined with block orders to obtain an optimal solution. 
Nevertheless, if we use less detailed, approximate cumulative demand and supply curves for each period and solve the clearing according to these, the MCPs resulting from this approximate clearing may be regarded as estimates of the 'real' prices, which would result from the original clearing problem, where the cumulative curves are considered in full detail.

The proposed approach consists of four steps.

\begin{enumerate}
    \item Starting from an initial bid set denoted by $\Phi^0$, composed of standard demand, standard supply bids and block orders, an other bid set $\Phi^1$ is derived, where the cardinality of standard demand and supply bids is less compared to $\Phi^0$, while the set of block orders is the same as in $\Phi^0$.
    The standard demand and supply bids of $\Phi^1$ are derived from the standard demand and supply bids of $\Phi^0$ via bid aggregation (BA). BA may be interpreted as using rough approximations of the cumulative demand and supply curves (derived from standard bids, so without block orders) in each period.
    \item In the second step, a standard market clearing problem is solved for the bid set $\Phi^1$.
    \item In the third step, based on the results of the clearing performed in step 2, estimates are derived for the possible ranges of the MCPs for each period.
    \item In the final step, the market clearing problem for the original bid set $\Phi^0$ is performed, taking into account the estimated MCP ranges as constraints derived in step 3. Such constraints already determine the acceptance and rejection of some standard bids, the variables of which may be removed from the optimization problem at the presolve stage, so the size of the problem is reduced.
\end{enumerate}

Let us note here that while for the aim of simplicity in this work we assume only 'step-wise' bids, with constant, quantity-independent prices, the proposed method may be also applied if linear bids are present (naturally, the objective function won't be linear anymore in this case as shown e.g. in \cite{cosmos2011}).

As we will see, it is possible that the second clearing problem in step 4 turns out to be infeasible because of the MCP constraints derived in step 3.
Furthermore, even if the clearing problem in step 4 is feasible, the final result of the approach is not necessarily an optimal solution of the original problem due to the potential acceptance or paradox rejection of block orders, which may be different in the clearings in steps 2 and 4. 
In fact, the success of the approach (in the context of  finding the 'real' dispatch, which may be obtained by solving the full scale clearing problem for the original bid set without any MCP constraints) depends on the bid aggregation pattern used in step 1, i.e. the same initial bid set with different aggregation patterns may result in proper or improper final outcome as well, depending on the actual aggregation pattern used. To overcome this potential obstacle, we propose to use multiple (in this paper 4) different aggregation patterns and run the proposed four step method for them in a parallel fashion. This approach, as we will see, ensures a high resulting success rate of the algorithm, regarding the final feasibility and welfare of the outcome, while simultaneously it is also able to significantly reduce the computational time in the case of problems with high number of standard bids.

The outline of the paper is as follows. Section \ref{sec_methods} describes the market model used in the study and introduces the 4 steps of the BA-based clearing. Section \ref{sec_pitfalls} uses simple examples to demonstrate the possible flaws of the approach, shows that the emergence of these issues depends on the aggregation pattern used and proposes a heuristic approach to derive 4 different aggregations in order to overcome these potential pitfalls.
Section \ref{sec_results} presents and discusses the results of numerical tests, while section \ref{sec_conclusions} concludes.

\section{Methods}
\label{sec_methods}

In this section we formally introduce the considered problem, describe its reference solution and then detail the various steps of the proposed approach.

\subsection{Formal description of the basic problem and reference solution}
\label{subsection_basic_model}

For the aim of simplicity, we consider a single-zone multiperiod market of $T$ periods in the proposed model ($t \in [1...T]$).
An order set denoted by $\Phi$
 consists of the set of standard demand bids ($SD$), the set of standard supply bids ($SS$) and the set of block orders ($BB$), which define the supply and demand for each market period.
In other words, $\Phi$ is an ordered triplet of the sets of various bid types considered in the model, i.e. $\Phi=(SD,SS,BB)$.

While a standard bid corresponds to a single trading period and may be partially accepted, block orders, which may include multiple trading periods and have the '\emph{fill-or-kill}' property, can be only entirely accepted or rejected in all included periods \cite{meeus2009block}.

Each standard bid $i \in SD$ or $\in SS$ is characterized by the relevant time period, its quantity ($Q_i$) and its price ($P_i$).
We assume that $Q_i>0$ in the case of demand bids (i.e. $i \in SD$) and $Q_i<0$ in the case of supply bids (i.e. $i \in SS$).
$SD_t$ and $SS_t$ denote the sets of standard demand and supply bids relevant for period $t$.
Without the loss of generality,
we assume that for each period standard bids are sorted according to bid price (descending in the case of demand and ascending in the case of supply), and their IDs are defined in accordance with this ordering (i.e. $(i<j) \rightarrow (P_j \leq P_i)$ if $i,j~\in~SD_t$, and $(i<j) \rightarrow (P_j \geq P_i)$ if $i,j~\in~SS_t$).
Each block bid $j$ is characterized by a start and end period denoted by $s_j$ and $e_j$ respectively, by a quantity $Q_j$ and by a bid price $P_j$. We assume that the latter two parameters are uniform for all periods included in the bid (in other words we do not consider 'profiled block bids' \cite{euphemia2015}).

Upper indices are used to distinguish between various bid sets.
The input of the considered market is the reference order set denoted by $\Phi^0=(SD^0,SS^0,BB^0)$.

\subsubsection{Reference solution: multi-period day-ahead market clearing problem formulation including block orders}
\label{subsec_general_formulation}

The task of the market clearing algorithm -- which is formulated as an optimization problem -- is to determine the sets of accepted and rejected bids and the respective MCPs for each period, which define the payoff values of accepted bids.

With respect to the formulation of the market clearing problem, we follow the approach of \cite{derinkuyu2015determination} in the sense that we explicitly include the MCPs as variables.
This will help us later to explicitly consider the estimated ranges, which limit the possible values of the MCPs in a simple manner by introducing inequality constraints for these variables.
The market model used in the proposed framework may be viewed as a simplification of the EUPHEMIA framework \cite{chatzigiannis2016european}, and has the following decision variables:  $x_i~\in~[0,1]~~i \in SD~\text{or}~\in SS$) denotes the acceptance indicator of the $i$-th standard bid, while $x_j~\in~\{0,1\}~~j \in BB$ denotes the acceptance variable of the $j$-th block bid.
$x$ denotes the vector of bid acceptance indicators.
The market clearing price of period $t$ is denoted by $MCP_t$, and $MCP\in \Real^{T}$ denotes the vector of the market clearing prices for all periods.
Problem 1 describes the optimization problem corresponding to the market clearing.

\paragraph*{Problem 1}

\begin{small}
\begin{align}
    &\max_{x,~MCP} \sum_t \left( \sum_{i\in SD^0_t\cup SS^0_t} x_i Q_i P_i + \sum_{j \in BB^0_t} x_j Q_j P_j \right) \label{eq_TSW_P1}\\
    \nonumber \\
    s.t.~~~~&\sum_{i \in SD^0_t\cup SS^0_t} x_i Q_i  + \sum_{j \in BB^0_t} x_j Q_j= 0 \qquad \forall\,t \label{eq_sup_dem_balance_P1} \\
    & x_i > 0 \rightarrow MCP_t \leq P_{i}, ~~~ x_{i} < 1 \rightarrow P_{i}\leq MCP_t \qquad \forall\,t\;\forall\, i\in SD^0_t\nonumber \\
&x_i > 0 \rightarrow P_{i} \leq  MCP_t, ~~~ x_{i} < 1 \rightarrow MCP_t \leq P_{i} \qquad \;\forall\,t\;\forall\, i \in SS^0_t \label{eq_BAS_P1}\\
&x_j > 0 \rightarrow (e^B_j-s^B_j+1)P_{j} \leq \sum_{s^B_j \leq t \leq e^B_j} MCP_t  \qquad \forall j\in BB^0~~~~,
\label{eq_BAB_P1}
\end{align}
\end{small}
where $BB^0_t$ is the set of block orders active in period $t$, namely $BB^0_t=\{j \in BB^0: s_j\le t\le e_j\}$

\paragraph*{Objective}
The objective function is to maximize the total social welfare (TSW) -- the 'total utility of consumption minus the total cost of production' -- over all periods, as described by (\ref{eq_TSW_P1}).

\paragraph*{Constraints}
The proposed simple clearing model considers on the one hand the power balance constraint, described by eq. (\ref{eq_sup_dem_balance_P1}), which applies for each period, and the bid acceptance constraints, which describe how the MCP is related to the set of accepted and rejected bids, described by eqs. (\ref{eq_BAS_P1}) and (\ref{eq_BAB_P1}) for simple and block bids respectively.
Regarding the bid acceptance constraints of block bids, constraint (\ref{eq_BAB_P1}) requires that the overall net payoff of the block bid must be non-negative.

While according to eq. (\ref{eq_BAS_P1}), the $MCP_t$ values explicitly determine the acceptance indicators of all simple bids for period $t$, for which $p^S_i \neq MCP_t$ (as zeros or ones), and allow partial acceptance ($x^S_i \in [0,1]$) for bids, for which $p^S_i = MCP_t$,
eq. (\ref{eq_BAB_P1}) shows that paradox rejection of block bids \cite{madani2014minimizing} is allowed (i.e. a block bid may be rejected, even if its net payoff is positive).
The implications may be formalized in the optimization framework e.g. by the so called 'bigM' method  \cite{bonami2015mathematical}, using auxiliary binary variables.

\paragraph*{Solution}

The solution of Problem 1 is denoted by $x^{ref}$ and $MCP^{ref}$, while the value of the objective function in the case of $x^{ref}$ and $MCP^{ref}$ is denoted by $TSW^{ref}$.

\subsection{Bid-aggregation based clearing}
\label{subsec_concept}

In this subsection, we describe the steps of the proposed alternative heuristic approach.

\subsubsection{Step 1: Aggregation of bids}
\label{subsec_nominal_aggregation}
The objective of step 1 is to obtain the aggregated bid set $\Phi^1=(SD^1,SS^1,BB^1)$ from the original bid set $\Phi^0=(SD^0,SS^0,BB^0)$. To this aim we introduce the concept of bid-aggregation (BA).
An aggregation operator, denoted by $\mathcal{A}: \Phi^0 \rightarrow \Phi^1$ assigns an aggregated bid set $\Phi^1$ to an input bid set $\Phi^0$. First of all let, us emphasize that block orders are not subject to the aggregation process, thus are unaffected by the aggregation operator (i.e. $BB^1=BB^0$).

Standard bids are clustered by price and aggregated separately for the demand and supply side for each market period.
$\mathcal{A}$ is defined via the vectors 
$\mathcal{A}_{D,t} \in \mathbb{N}^{|SD_t|}$ and $\mathcal{A}_{S,t} \in \mathbb{N}^{|SS_t|}$  ($t \in \{1,\ldots,T\}$), where
$|SD_t|$ and $|SS_t|$ denote the cardinality of the sets $SD_t$ and $SS_t$ respectively, i.e. the number of demand and supply bids relevant for period $t$ (and $\mathbb{N}$ is the set of natural numbers). $\mathcal{A}_{D,t}(i)$ equals to the index of the aggregated demand bid, in which the $i$-th standard demand bid of period $t$ is aggregated (similarly for $\mathcal{A}_{S,t}(i)$). 
The index set of aggregated (standard) demand and supply bids regarding period $t$ is denoted by $ASD_t$ and $ASS_t$ respectively. Since (as we will see) consecutive bids are aggregated, by construction $ASD_t$ and $ASS_t$ are increasingly/decreasingly sorted by bid price. In addition, we assume that the aggregation method ensures that standard demand or supply bids with the same price for the same period are always aggregated into the same aggregated demand or supply bid.

\paragraph*{The aggregation rule for \texorpdfstring{$n$}{n} bids}
We describe the aggregation of standard supply bids, but the method is the same for demand bids as well.
Let as assume that we have $n$ standard supply bids for a given time period $t$, with consecutive indices ranging from $i$ to $i+(n-1)$ $\in SS_t$, which are aggregated by the aggregation  ($\mathcal{A}$) into a single (aggregated) bid of index $m$, i.e. 
$\mathcal{A}_{S,t}(i)=\mathcal{A}_{S,t}(i+1)= \ldots = \mathcal{A}_{S,t}(i+(n-1))=m$.

In this case the parameters of the resulting aggregated bid by definition will be computed as described by eq. (\ref{eq:aggregation}), which shows that the bid price ($P_{m}$) of the aggregated supply bid $m$, is computed from the quantity-weighted average of original bid prices, while its quantity ($Q_{m}$) will be equal to the total quantity of the bids included.
\begin{equation}
\begin{aligned}
&P_{m}= \frac{\sum_{k = i}^{i+(n-1)} P_k Q_k}{\sum_{k = i}^{i+(n-1)} Q_k} ~~~~Q_{m} = \sum_{k = i}^{i+(n-1)} Q_k  \label{eq:aggregation}
\end{aligned}
\end{equation}

\paragraph*{Example aggregation}

\begin{figure}[h!]
    \centering
    \includegraphics[scale=0.7]{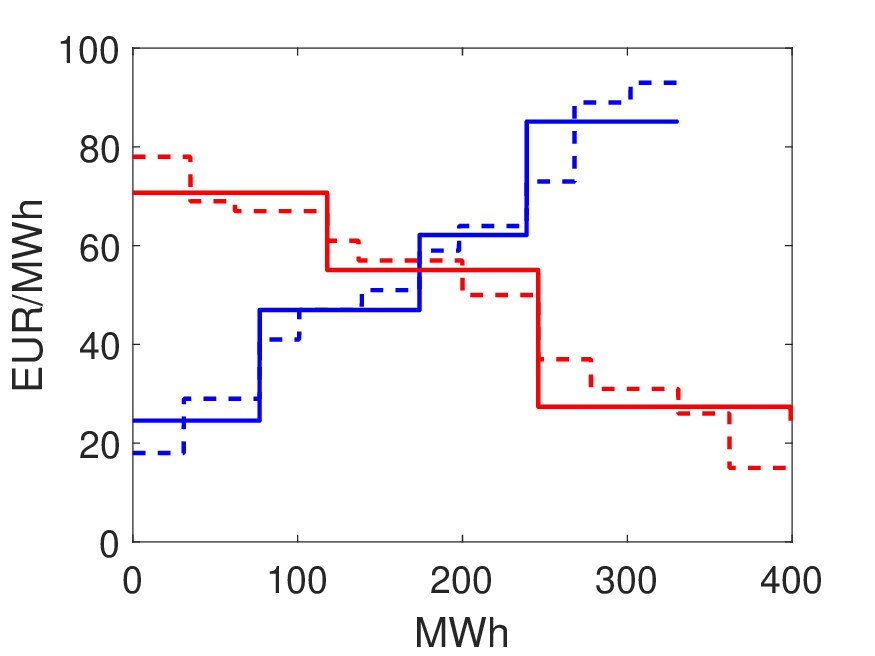}
    \caption{Cumulative demand and supply curves of the original ($\Phi^0$) and of the aggregated ($\Phi^1$) bid set, described in Tables \ref{tab_bid_data_demo_1} and \ref{tab_bid_data_demo_res_1} and denoted by dashed and continuous lines respectively.}
    \label{fig_aggreg_ex}
\end{figure}

To provide a simple example, let us consider the 1-period bid set $\Phi^0$ containing only standard bids (block orders are not affected by the aggregation) described in Table \ref{tab_bid_data_demo_1}, and the aggregation pattern described by eq. (\ref{eq_example_aggr_pattern}). In this case the aggregation will result in the bid set $\Phi^1$ described in Table \ref{tab_bid_data_demo_res_1}.
The cumulative demand and supply curves of the original and of the aggregated bid set are depicted in Fig. \ref{fig_aggreg_ex}.

\begin{table}[h!]
\begin{small}
    \centering
    \begin{tabular}{|c|c|c|c|c|c|}
    \hline
    ID  & $Q$ [MWh] & $P$ [EUR/MWh] & ID  & $Q$ [MWh]& $P$ [EUR/MWh]\\ \hline
     1 &  35 &  78 & 11 &-31 & 18  \\
     2 &  27 &  69 & 12 &-46 & 29  \\
     3 &  56 &  67 & 13 &-24 & 41  \\
     4 &  19 &  61 & 14 &-38 & 47  \\
     5 &  63 &  57 & 15 &-35 & 51  \\
     6 &  46 &  50 & 16 &-24 & 59  \\
     7 &  32 &  37 & 17 &-41 & 64  \\
     8 &  53 &  31 & 18 &-29 & 73  \\
     9 &  31 &  26 & 19 &-34 & 89  \\
    10 &  37 &  15 & 20 &-28 & 93  \\
    \hline
    \end{tabular}
    \label{tab_bid_data_demo}
    \caption{Example original bid data: Standard bids ($SD^0$ and $SS^0$) of $\Phi^0$.}
    \label{tab_bid_data_demo_1}
    \end{small}
\end{table}

\begin{align}
    \mathcal{A}_{D,1}=\left( \begin{array}{c} 
    1\\
    1\\
    1\\
    2\\
    2\\
    2\\    
    3\\
    3\\
    3\\
    3\\
    \end{array} \right)~~~
    \mathcal{A}_{S,1}=\left( \begin{array}{c} 
    4\\
    4\\
    5\\
    5\\
    5\\    
    6\\
    6\\
    7\\
    7\\
    7\\
    \end{array} \right)~~~
    \label{eq_example_aggr_pattern}
\end{align}

\begin{table}[h!]
\begin{small}
    \centering
    \begin{tabular}{|c|c|c|c|c|c|}
    \hline
    ID &  $Q$ [MWh] & $P$ [EUR/MWh] & ID & $Q$ [MWh]& $P$ [EUR/MWh]\\ \hline
    A1 &  118 & 70.7 &  A4 & -77 & 24.6  \\
    A2 &  128 & 55.1 &  A5 & -97 &   47  \\
    A3 &  153 & 27.4 &  A6 & -65 & 62.2  \\
       &      &      &  A7 & -91 & 85.1  \\
    \hline
    \end{tabular}
    \caption{Resulting aggregated bid data: Standard bids of $\Phi^1$ ($SD^1$ and $SS^1$).}
    \label{tab_bid_data_demo_res_1}
    \end{small}
\end{table}

\paragraph*{Nominal aggregation}
The most simple approach is to form clusters of bids according to their price (using a well-defined clustering approach) and aggregate the components of these clusters into a single bid.
This approach will be called the \emph{nominal aggregation} and will be denoted by $\mathcal{A}^N$.
As we will see later, based on this reference aggregation, other aggregations described by different aggregation patterns will be defined as well.

The clustering method used in the current work (see \nameref{Appendix A}) is performed by the  \textit{cluster} function of \textit{MATLAB} \cite{Matlab2019}, which takes three inputs: the linkage information of different clusters, the method it uses to determine the way to separate the clusters 
and the value of the method's parameter. 
The clustering defines the vectors $\mathcal{A}^N_{S,t}$ and $\mathcal{A}^N_{D,t}$ (for each time period $t$). Naturally, any other clustering approach may be also used for this aim.


\subsubsection{Step 2: Clearing of the aggregated bids}

In step 2, a standard market clearing problem with identical form to Problem 1 is solved, considering the the aggregated bid set $\Phi^1$, which means that one must substitute $SD^0_t$ with $SD^1_t$ and $SS^0_t$ with $SS^1_t$ in the expressions ($BB^1=BB^0$, so $BB^1_t=BB^0_t$ and no substitution is needed).

\subsubsection{Step 3: Derivation of estimated MCP ranges}
\label{subsec_step_3}

Solving the market clearing problem based on the aggregated bids ($\Phi^1$) will of course (typically) not provide the MCPs equal to $MCP^{ref}$, but after solving the clearing for the aggregated bids, assumptions regarding the possible range of MCP values for each period are derived. 




Regarding the solution of Problem 1 for $\Phi^1$, obtained in step 2, three possibilities may arise for each period $t$
 \begin{itemize}
   \item One of the aggregated demand bids (the price setter bid) is partially accepted.
   \item One of the aggregated supply bids is partially accepted.
   \item The acceptance indicator of all aggregated bids is 0 or 1.
 \end{itemize}

If there are bids with identical price on the supply and demand side for period $t$, and these turn out to be the price setters (their price is equal to $MCP_t$) -- i.e. the intersection point of the cumulative curves is not unique --, it may happen that in the particular solution there are more partially accepted bids for period $t$. In this case however, a solution with identical $MCP_t$ and $TSW$ exists, where only one of these bids is partially accepted, and the traded quantity is higher -- in this case we consider this solution. Furthermore, let us note that since bids of $\Phi^0$ with the same price for any period are always aggregated into a single bid of $\Phi^1$, there are no standard bids of the same price in $\Phi^1$ for either period for the same side (demand or supply), so it may not happen that multiple bids of one side are price setter bids (in this case, these could be partially accepted in multiple configurations, as long as their overall quantity is equal). 

Let us suppose that there is an aggregated bid on the demand side that has been partially accepted -- the price setter bid (denote its index with $m_{APSD}$), and on the supply side there is at least one fully accepted and one fully rejected aggregated bid. In this case, we can identify the two closest aggregated bids (close in the sense of bid price): the fully accepted aggregated supply bid with the highest price ($m_{AS}$), and the fully rejected aggregated supply bid with the lowest price ($m_{RS}$).

The estimated $MCP$ interval for period $t$ based on the outcome of step 1 may be obtained as described by eq. (\ref{eq_MCP_est_bounds}).
The interpretation of eq. (\ref{eq_MCP_est_bounds}) is that the assumed upper and lower bounds for the $MCP$ regarding period $t$ (denoted by $\overline{MCP}_t$ and $\underline{MCP}_t$ respectively) are derived based on the bid prices of original (not-aggregated) bids, which have been included in the 3 distinguished aggregated bids ($m_{APSD},~m_{AS}$ and $m_{RS}$).

\begin{align}
& \overline{MCP}_t= max \left(
max (P_i) ~~(i: \left( \mathcal{A}_{D,t}(i)=m_{APSD} \right) ),~min (P_i)~~(i: \left( \mathcal{A}_{S,t}(i)=m_{RS} \right))
\right ) \nonumber\\
& \underline{MCP}_t=min \left( 
min (P_i) ~~(i: \left( \mathcal{A}_{D,t}(i)=m_{APSD} \right) ),~
max (P_i)~~(i: \left( \mathcal{A}_{S,t}(i)=m_{AS} \right) )
\right)
\label{eq_MCP_est_bounds}
\end{align}
 where $i: \left( \mathcal{A}_{S,t}(i)=m \right)$ denotes the set of original standard supply bids, which have been aggregated into the aggregated supply bid of index $m$, according to the aggregation $\mathcal{A}$, and $\left( \mathcal{A}_{D,t}(i)=m \right)$ denotes the set of original standard demand bids, which have been aggregated into the aggregated demand bid of index $m$, according to the aggregation $\mathcal{A}$.
 
\begin{figure}[h!]
    \centering
    \includegraphics[scale=0.7]{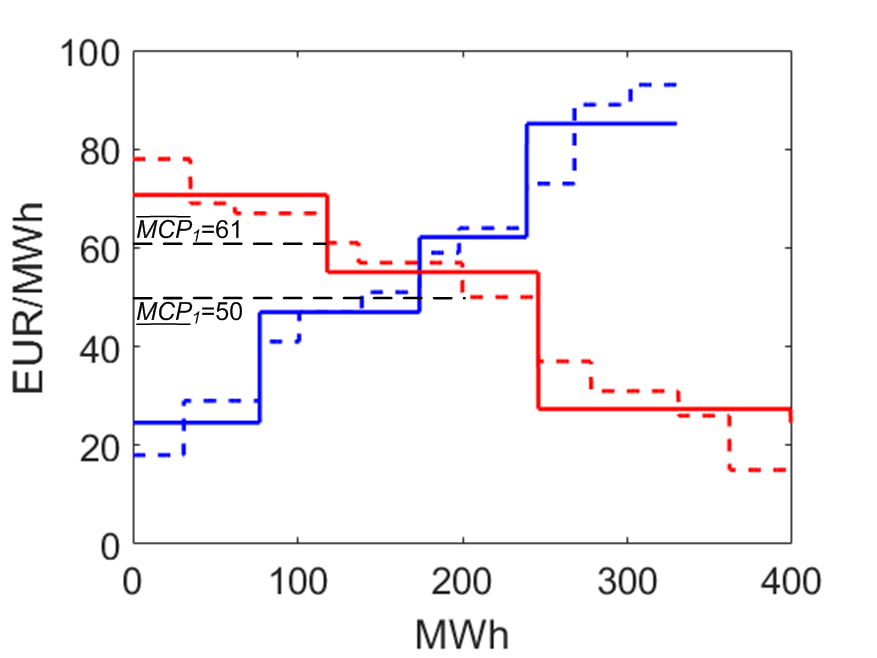}
    \caption{Example for the derivation of $\overline{MCP}_t$ and $\underline{MCP}_t$}
    \label{fig_MCP_deriv_ex}
\end{figure}

To give an example, let us consider the aggregation described in subsection \ref{subsec_nominal_aggregation}, where the original bid set $\Phi^0$ and the aggregated bid set $\Phi^1$ are described in Tables \ref{tab_bid_data_demo_1} and
\ref{tab_bid_data_demo_res_1}.
In this case, the price setter bid is the aggregated demand bid A2. $\overline{MCP}_1$ may be derived as taking the maximum of (a) the maximum of the prices of the components of A2 (61) and (b) the minimum price of the components of the first rejected supply bid, A6 (59), which results in $\overline{MCP}_1=61$. Similarly $\underline{MCP}_1$ may be derived as taking the minimum of (a) the minimum of the prices of the components of A2 (50) and (b) the maximum price of the components of the last accepted supply bid, A5 (51), which results in $\underline{MCP}_1=50$.
An example depicting the derivation of $\overline{MCP}_t$ and $\underline{MCP}_t$ is depicted in Fig. \ref{fig_MCP_deriv_ex}.

Let us note that in this particular case the price setter bid was a demand bid both in $\Phi^0$ and also in $\Phi^1$, however in general, this is not the case -- see the further examples detailed later in the paper.

If there is an aggregated bid on the supply side that has been partially accepted, and there is at least one fully accepted and one fully rejected aggregated demand bid, we can similarly identify the two closest aggregated bids on the demand side, and use the same approach as described by eq. (\ref{eq_MCP_est_bounds_2}), where $m_{APSS}$ denotes the index of the partially accepted (price setter) supply bid, $m_{AD}$ denotes the index of the fully accepted aggregated demand bid with with the lowest price and $m_{RD}$ stands for the index of the fully rejected aggregated demand bid with the highest price.

\begin{align}
& \overline{MCP}_t= max \left(
max (P_i) ~~(i: \left( \mathcal{A}_{S,t}(i)=m_{APSS} \right) ),~min (P_i)~~(i: \left( \mathcal{A}_{D,t}(i)=m_{AD} \right))
\right ) \nonumber\\
& \underline{MCP}_t=min \left( 
min (P_i) ~~(i: \left( \mathcal{A}_{S,t}(i)=m_{APSS} \right) ),~
max (P_i)~~(i: \left( \mathcal{A}_{D,t}(i)=m_{RD} \right) )
\right)
\label{eq_MCP_est_bounds_2}
\end{align}

\paragraph*{Remark 1\\}
It may happen that there is no partially accepted aggregated bid in the result of the aggregated clearing. In this case, the component bids of the fully accepted aggregated supply bid with the highest price ($m_{AS}$), and the fully accepted aggregated demand bid with the lowest price ($m_{AD}$) may be used as reference. On these terms, $\overline{MCP}_t$ and $\underline{MCP}_t$ may be obtained as described by eq. (\ref{eq_MCP_est_bounds_3}).

\begin{align}
& \overline{MCP}_t= 
max (P_i) ~~(i: \left( \mathcal{A}_{D,t}(i)=m_{AD} \right) ))
 \nonumber\\
& \underline{MCP}_t=min (P_i) ~~(i: \left( \mathcal{A}_{S,t}(i)=m_{AS} \right) ))
\label{eq_MCP_est_bounds_3}
\end{align}

Some further cases must be also considered.

\begin{itemize}

\item In the case when the price setter bid is a partially accepted demand (supply) bid, it may happen that no fully accepted supply (demand) bid is 
present. In this situation we omit the respective value in the calculations, and $\overline{MCP}_t$ is determined by the maximum of prices among the component bids of the price setter aggregated bid.

\item If the price setter bid is a partially accepted demand bid, it may happen that there is no fully accepted aggregated supply bid, only an accepted block bid. In this case, the price of the block bid is used in the calculations.

\item It may also happen that there is no partially accepted aggregated bid, furthermore there are no accepted aggregated bids at all. This is possible only if the price of the lowest price supply bid (aggregated or block) exceeds the price of the highest price aggregated demand bid. In this case these, the components of these two bids may be used as reference.

\end{itemize}

\paragraph*{Remark 2\\}

Let us note that in this paper we proposed an approach for this estimation based on the bid price of 'component' bids of fully accepted and rejected aggregated bids close to the price setter bid, but other estimations are also possible.
There is a natural trade-off in the estimation process: Narrower derived $MCP$ intervals will imply more computational gain in step 4, but are also more likely to result in infeasible problems (see the latter example).

\subsubsection{Step 4: Clearing of the original bids according to the derived MCP estimates}
\label{subsec_step_4}

In step 4 we solve the clearing problem for the original bid set $\Phi^0$, while also consider the constraints corresponding to $\overline{MCP}_t$ and $\underline{MCP}_t$ values derived in subsection \ref{subsec_step_3}. Problem 2 (which is essentially Problem 1 complemented with the MCP constraints) describes the optimization problem of the clearing.

\paragraph*{Problem 2}

\begin{small}
\begin{align}
    &\max_{x,~MCP} \sum_t \left( \sum_{i\in SD^0_t\cup SS^0_t} x_i Q_i P_i + \sum_{j \in BB^0_t} x_j Q_j P_j \right) \label{eq_TSW_P2}\\
    \nonumber \\
    s.t.~~~~&\sum_{i \in SD^0_t\cup SS^0_t} x_i Q_i  + \sum_{j \in BB^0_t} x_j Q_j= 0 \qquad \forall\,t \label{eq_sup_dem_balance_P2} \\
    & x_i > 0 \rightarrow MCP_t \leq P_{i}, ~~~ x_{i} < 1 \rightarrow P_{i}\leq MCP_t \qquad \forall\,t\;\forall\, i\in SD^0_t\nonumber \\
&x_i > 0 \rightarrow P_{i} \leq  MCP_t, ~~~ x_{i} < 1 \rightarrow MCP_t \leq P_{i} \qquad \;\forall\,t\;\forall\, i \in SS^0_t \label{eq_BAS_P2}\\
&x_j > 0 \rightarrow (e^B_j-s^B_j+1)P_{j} \leq \sum_{s^B_j \leq t \leq e^B_j} MCP_t  \qquad \forall j\in BB^0
\label{eq_BAB_P2}\\
& \underline{MCP}_t \leq MCP_t \leq \overline{MCP}_t \qquad \forall t\label{eq_MCP_con_P2}
\end{align}
\end{small}

\paragraph*{Remark:  Presolve of Problem 2}

The MCP constraints described by (\ref{eq_MCP_con_P2})
explicitly determine the acceptance values for a subset of original bids.
According to the bid acceptance conditions of standard bids described by (\ref{eq_BAS_P2}), the constraint (\ref{eq_MCP_con_P2}) will force a subset of standard bids to be fully accepted or fully rejected.
Regarding the block orders, the derived $\overline{MCP}_t$ and $\underline{MCP}_t$ values may imply  rejection for a subset of them (MCP bounds can not imply acceptance for block orders, since paradox rejection is allowed).
The acceptance indicator variables of such bids (whose acceptance is explicitly determined by the assumed $MCP_t$ ranges) may be removed from further computations thus the effective problem size is reduced.

Due to the bids with fixed acceptance indicators in most cases a power imbalance is implied for each period. In order to get a final result which respects the power balance constraint, this imbalance must be taken into account and the remaining bids must be cleared in a way which compensates for this imbalance (for each period). 

Let us furthermore note that in contrast to Problem 1, Problem 2 is not necessary feasible. 

\section{Potential pitfalls of the proposed approach and how to avoid them}
\label{sec_pitfalls}
In this section, we first introduce two examples highlighting the potential flaws of the presented approach and propose a method, which helps to avoid such issues.

\subsection{Example I - Suboptimal result}
\label{subsec_example}
In this subsection we present a simple, one-period example to demonstrate the operation of the BA based two-stage clearing, and show that the approach may result in a suboptimal solution.

Table \ref{tab_bid_data_ex_subopt} summarizes $\Phi^0$ of example I. The prefix 'B' (in B1) stands for a block order (or 'fill-or-kill' bid), while the other bids are standard bids.

\begin{table}[h!]
\begin{small}
    \centering
    \begin{tabular}{|c|c|c|c|c|c|}
    \hline
    ID & $Q$ [MWh] & $P$ [EUR/MWh] & ID & $Q$ [MWh]& $P$ [EUR/MWh]\\ \hline
     1 &       154 &       104 &        8 &       -121 &      23.9  \\
     2 &       104 &        89 &         9 &      -84.4 &      26.6  \\
     3 &        65 &        83  &      10 &      -48.9 &        52 \\
     4 &        51 &        56&        11 &       -55 &      62.7   \\
     5 &        99 &        49 &       12 &     -50.6 &      76.8  \\
     6 &        52 &        46 &       13 &     -73.4 &      85.2  \\
     7 &        36 &        34 & B1 &   -150 &        50  \\

    \hline
    \end{tabular}
    \caption{The bid set $\Phi^0$ of example I}
    \label{tab_bid_data_ex_subopt}
    \end{small}
\end{table}

\subsubsection{Reference solution of example I}

\begin{figure}[h!]
    \centering
\includegraphics[width=8cm]{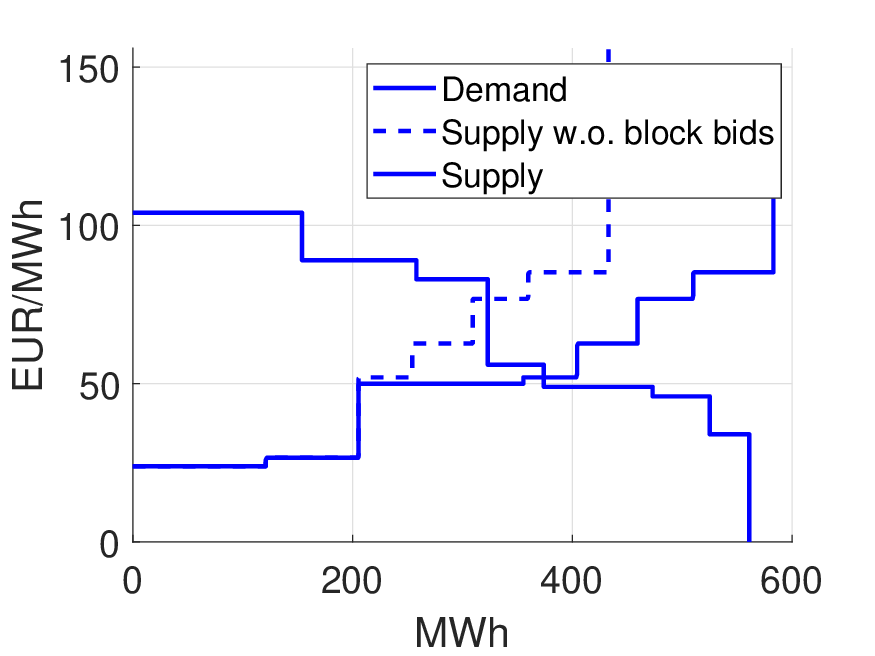}
    \caption{Cumulative supply and demand curves implied by the bids included in table \ref{tab_bid_data_ex_subopt}.}
    \label{subopt_EX_1}
\end{figure}

Figure \ref{subopt_EX_1} depicts the cumulative supply and demand curves implied by the bids included in Table \ref{tab_bid_data_ex_subopt}, and highlights that the conventional market clearing problem described by eqs. (\ref{eq_TSW_P1} - \ref{eq_BAB_P1}) results in $MCP^{ref}=52$ EUR, and the acceptance of the block bid $B1$. $TSW^{ref}$ is equal to 19919 EUR in this case.

\subsubsection{BA-based clearing of example I}

\begin{figure}[h!]
    \centering
\includegraphics[width=8cm]{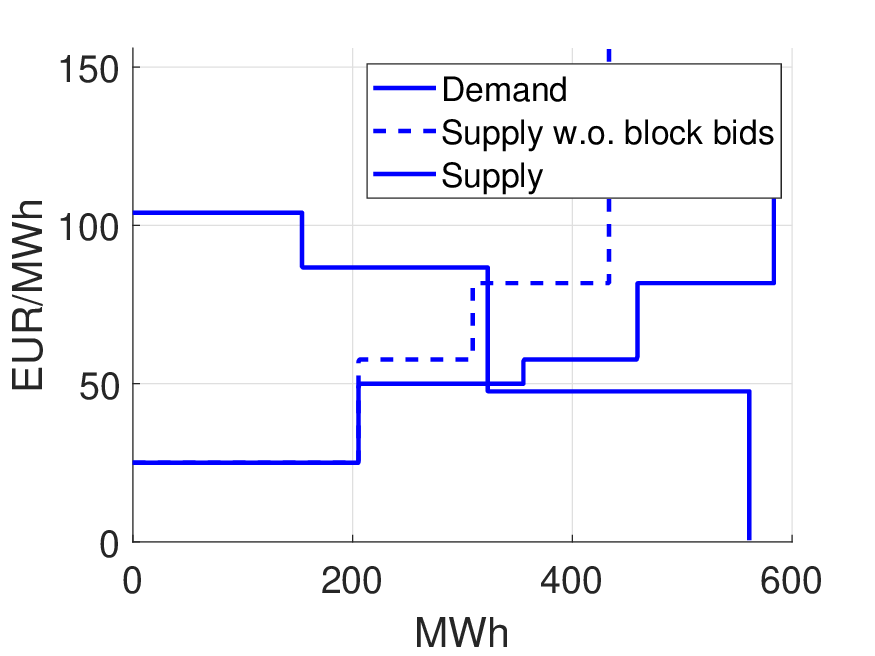}
    \caption{Cumulative supply and demand curves of $\Phi^1$ of example I.}
    \label{subopt_EX_1_S1}
\end{figure}

The clustering of bids according to prices (as described in the case of nominal aggregation) results the aggregation vectors $\mathcal{A}^N_D$ and $\mathcal{A}^N_S$ -- since the example holds only one period we omit the lower index $t$ for $\mathcal{A}_{D,t}$ and $\mathcal{A}_{S,t}$ -- summarized in Table \ref{tab_ex_1_subopt_aggr_patt}, and, according to eq. (\ref{eq:aggregation}), the aggregated bid set ($\Phi^1$) described in Table \ref{tab_bid_data_ex_1_subopt_aggr}. 
The cumulative supply and demand curves of 
$\Phi^1$ are depicted in Figure \ref{subopt_EX_1_S1}.

\begin{table}[h!]
\begin{small}
    \centering
    \begin{tabular}{|c|c|c|c|}
    \hline
    ID & $\mathcal{A}^N_D(i)$ & ID & $\mathcal{A}^N_S(i)$ \\ \hline
     1 &         1 &        8 &         4  \\
     2 &         2 &        9 &         4  \\
     3 &         2 &        10 &        5  \\
     4 &         3 &        11 &        5  \\
     5 &         3 &        12 &        6  \\
     6 &         3 &        13 &        6  \\
     7 &         3 &          &            \\
    \hline
    \end{tabular}
    \caption{Vectors $\mathcal{A}^N_D$ and $\mathcal{A}^N_S$ describing the nominal aggregation pattern $\mathcal{A}^N$ for example I}
    \label{tab_ex_1_subopt_aggr_patt}
    \end{small}
\end{table}

\begin{table}[h!]
\begin{small}
    \centering
    \begin{tabular}{|c|c|c|c|c|c|}
    \hline
    ID & $Q$ [MWh] & $P$ [EUR/MWh] & ID & $Q$ [MWh]& $P$ [EUR/MWh]\\ \hline
     A1 &       154  &   104    &        A4 &  -205.4 &  25.01 \\
     A2 &       169 &   86.69   &        A5 &  -103.9 &  57.66 \\
     A3 &       238  & 47.58    &      A6 &    -124. &  81.77  \\
       &          &           &  B1 &   -150 &        50   \\
    \hline
    \end{tabular}
    \caption{Bid data of aggregated bids ($\Phi^1$) in the case of example I and nominal aggregation}
    \label{tab_bid_data_ex_1_subopt_aggr}
    \end{small}
\end{table}

As Figure \ref{subopt_EX_1_S1} shows,
if we perform step 2 (the clearing of the aggregated bids), we see that the price setter bid will be the aggregated supply bid A6 ($m_{APSS}=6$ -- see the dashed supply curve), and the block bid will be paradoxically rejected in this case.

Regarding step 3 and the derivation of the MCP estimates, in this particular case, the prices of the original bids aggregated into the price setter bid A6 (12 and 13 in $\Phi^0$) are 76.8 and 85.2 respectively.
Next, we have to consider the fully accepted aggregated demand bid with the lowest price, which corresponds to the ID A2 ($m_{AD}=2$) in the aggregated bid set. As Table \ref{tab_ex_1_subopt_aggr_patt} shows, bids 2 and 3 (according to original bid IDs) have been aggregated into this bid, with prices 89 and 83.
In addition, we have to look for the fully rejected aggregated demand bid with the highest price, which is A3 ($m_{RD}=3$) in $\Phi^1$. As Table \ref{tab_ex_1_subopt_aggr_patt} shows, bids 4-7 (regarding the original IDs in $\Phi^0$) have been aggregated into this bid, with prices 56, 49, 46 and 34 respectively.

According to eq. (\ref{eq_MCP_est_bounds_2}),
\begin{align*}
    \overline{MCP}&=\max (\max \{ 76.8,~85.2\},~ \min \{89,~83 \})=85.2 \\
    \underline{MCP}&=\min (\min \{ 76.8,~85.2\},~\max \{ 56,~ 49, ~46,~34\}=56~~. 
\end{align*}

The derived $MCP$ constraints imply full acceptance or rejection for a subset of original bids as described in Table \ref{table:ex_1_fixed_AIs}.

\begin{table}[h!]
\begin{small}
    \centering
    \begin{tabular}{|c|c|c|c|}
    \hline
    ID & $x_i$ & ID & $x_i$ \\ \hline
     1 &         1 &        8 &         1  \\
     2 &         1 &        9 &         1  \\
     3 &         - &        10 &         1  \\
     4 &         - &        11 &         -  \\
     5 &         0 &        12 &         -  \\
     6 &         0 &        13 &         -  \\
     7 &         0 &        B1 &         -  \\
\hline
    \end{tabular}
    \caption{Fixed and undetermined acceptance indicators of original bids after applying the MCP constraints.}
    \label{table:ex_1_fixed_AIs}
    \end{small}
\end{table}

The fixed acceptance indicators imply 258 units of accepted demand and 254.3 units of accepted supply, thus the second clearing step for the remaining undetermined bids must be performed in order to account for this imbalance, i.e. to obtain a surplus of 3.7 units on the supply side. The solution of Problem 2 in step 4 with the derived MCP constraints will determine the remaining (not yet fixed) bid acceptance indicators as summarized in Table \ref{table:Ex_1_final_AIs}, and resulting in the $MCP$ of 76.8 EUR. It is important to note that in contrast to the reference solution, the block bid B1 is rejected in this case.
The resulting $TSW$ equals to 18487 EUR instead of $TSW^{ref}=19919$ thus a suboptimal solution has been obtained by the BA based clearing.

\begin{table}[h!]
\begin{small}
    \centering
    \begin{tabular}{|c|c|c|c|c|c|}
    \hline
    ID & $x^{ref}_i$ & $x^{BA}_i$ & ID & $x^{ref}_i$ & $x^{BA}_i$\\ \hline
    1 &       1 &         1 &        8 &         1 &          1  \\
    2 &       1 &         1 &        9 &         1 &          1  \\
    3 &       1 &         1 &        10 &        0.38 &       1  \\
    4 &       1 &         0 &        11 &         0 &         1  \\
    5 &       0 &         0 &        12 &         0 &         0.27  \\
    6 &       0 &         0 &        13 &         0 &         0  \\
    7 &       0 &         0 &        14 &         1 &         0  \\
    \hline
    \end{tabular}
    \caption{Final acceptance indicators of original bids in the case of the BA based method ($x^{BA}_i$) compared to the original reference solution $x^{ref}_i$ (example I).}
    \label{table:Ex_1_final_AIs}
    \end{small}
\end{table}

\subsection{Example II - Infeasibility}
\label{subsec_example_2}
To show that suboptimality is not the only issue which may arise in the process of the BA based clearing, in this subsection we present an additional example to demonstrate that Problem 2 of the proposed method may prove to be infeasible as well.

Table \ref{tab_bid_data_infeas_ex} summarizes the bid parameters ($\Phi^0$) of example II. B1 is a block bid, while the other bids are standard bids.

\begin{table}[h!]
\begin{small}
    \centering
    \begin{tabular}{|c|c|c|c|c|c|}
    \hline
    ID & $Q$ [MWh] & $P$ [EUR/MWh] & ID & $Q$ [MWh]& $P$ [EUR/MWh]\\ \hline
   1 &          130 &        100 &        8 &       -160 &        20  \\
     2 &        100 &        90 &        9 &       -80 &        30  \\
     3 &        50 &        80 &        10 &       -50 &        52  \\
     4 &        100 &        70 &        11 &       -60 &        53  \\
     5 &        50 &        48 &        12 &       -60 &       72  \\
     6 &        50 &        42 &        13 &       -70 &       83  \\
     7 &        40 &        30 &    B1 &       -150 &       50  \\
    \hline
    \end{tabular}
    \caption{$\Phi^0$ of example II}
    \label{tab_bid_data_infeas_ex}
    \end{small}
\end{table}

\subsubsection{Reference solution for example II}

\begin{figure}[h!]
    \centering
\includegraphics[width=8cm]{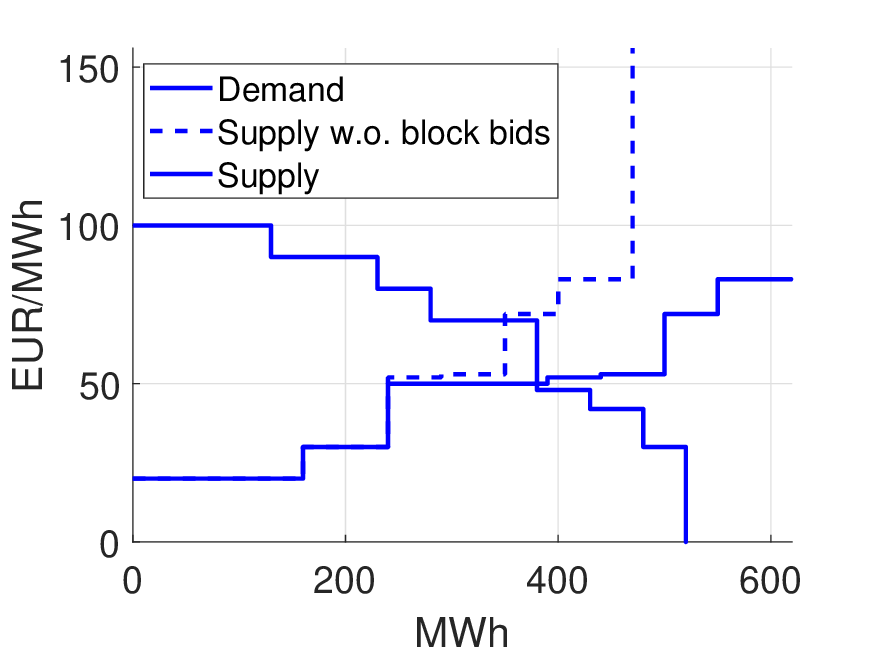}
    \caption{Cumulative supply and demand curves implied by the bids included in Table \ref{tab_bid_data_infeas_ex}.}
    \label{infeas_EX_1}
\end{figure}

Figure \ref{infeas_EX_1} depicts the cumulative supply and demand curves implied by the bids included in Table \ref{tab_bid_data_infeas_ex}, and shows that the conventional market clearing problem results in $MCP^{ref}$ of 70 EUR, and the paradox rejection of B1.

\subsubsection{BA-based solution for example II}

\begin{figure}[h!]
    \centering
\includegraphics[width=8cm]{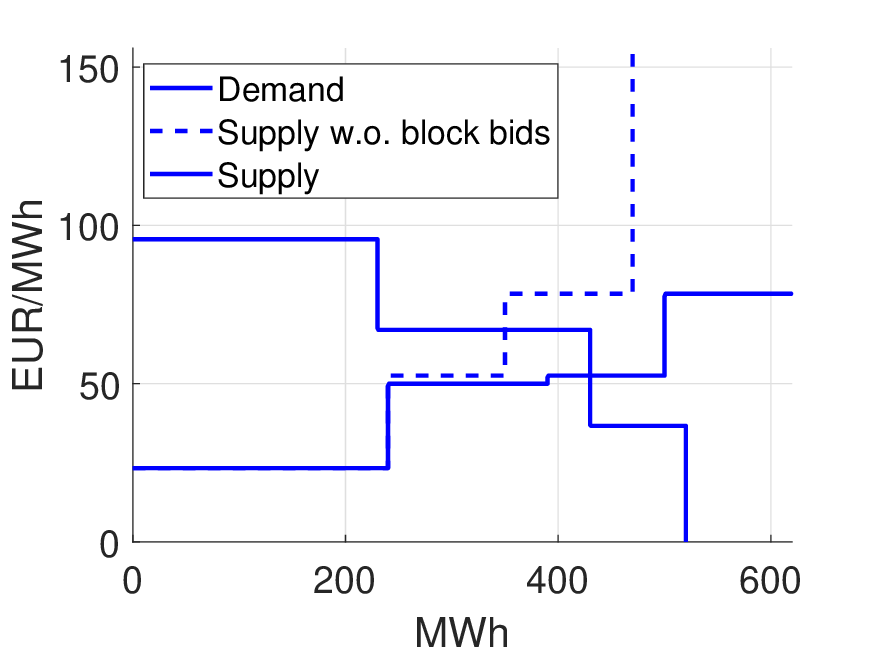}
    \caption{Cumulative supply and demand curves resulting from $\Phi^1$ defined in Table \ref{tab_infeas_aggr_patt}}
    \label{fig:infeas_EX_1_aggr}
\end{figure}

Let us consider the aggregation described in Table \ref{tab_infeas_aggr_patt}, which will result in the bid set $\Phi^1$ summarized in Table \ref{tab_bid_data_aggr_infeas_ex}, and in the cumulative supply and demand curves for step 2 (the clearing of the aggregated bids) depicted in Fig \ref{fig:infeas_EX_1_aggr}.

\begin{table}[h!]
\begin{small}
    \centering
    \begin{tabular}{|c|c|c|c|}
    \hline
    ID & $\mathcal{A}^N_D(i)$ & ID & $\mathcal{A}^N_S(i)$ \\ \hline
     1 &         1 &        8 &         4  \\
     2 &         1 &        9 &         4  \\
     3 &         2 &        10 &         5  \\
     4 &         2 &        11 &         5  \\
     5 &         2 &        12 &         6  \\
     6 &         3 &        13 &         7  \\
     7 &         3 &         &           \\
    \hline
    \end{tabular}
    \caption{The aggregation pattern of the nominal aggregation ($\mathcal{A}^N$) for example II}
    \label{tab_infeas_aggr_patt}
    \end{small}
\end{table}

\begin{table}[h!]
\begin{small}
    \centering
    \begin{tabular}{|c|c|c|c|c|c|}
    \hline
    ID & $Q$ [MWh] & $P$ [EUR/MWh] & ID & $Q$ [MWh]& $P$ [EUR/MWh]\\ \hline
     A1 &       230 &     95.65 &        A4 &  -240 &   23.33 \\
     A2 &       200 &        67 &        A5 &  -110 &   52.55 \\
     A3 &       90  &     36.67 &       A6 &   -120  &  78.42  \\
       &          &          &  B1& -150 &  50    \\
    \hline
    \end{tabular}
    \caption{Bid data of aggregated bids ($\Phi^1$) in the case of example II and nominal aggregation}
\label{tab_bid_data_aggr_infeas_ex}
    \end{small}
\end{table}

If we perform the clearing of the aggregated bids (step 2), we see that the price setter bid will be the aggregated supply bid A5, with $P=52.55$. In contrast to the nominal solution for $\Phi^0$, the block bid B1 will be accepted in this case.

In this case, the prices of the original bids aggregated into the price setter bid (IDs 10 and 11 in the original setting) are 52 and 53.
Next, we have to look for the fully accepted aggregated demand bid with the lowest price, which corresponds to the ID A2 ($m_{AD}=2$) in the aggregated bid set. As Table \ref{tab_infeas_aggr_patt} shows, bids 3,4 and 5 (regarding the original IDs) have been aggregated into this bid, among which the minimum price is 48.
In addition, we have to look for the fully rejected aggregated demand bid with the highest price, which is A3 ($m_{RS}=3$) in the aggregated bid set. As Table \ref{tab_infeas_aggr_patt} shows, bids 6 and 7 (regarding the original IDs) have been aggregated into this bid, among which the maximum price is 42.

According to eq. (\ref{eq_MCP_est_bounds_2}), the $MCP$ constraints may be derived as

\begin{align*}
& \overline{MCP}_t= \max \left( \max \{ 52,~53\} ), \min 
\{80,~70,~48 \}\right)=53 \\
& \underline{MCP}_t= \min \left( 
\min \{ 52,~53\},~\max \{ 42, 30 \} \right)=42~~.
\end{align*}

The $MCP$ constraints imply full acceptance or rejection for a subset of the original bids as described in Table \ref{table:ex_2_fixed_AIs}.

\begin{table}[h!]
\begin{small}
    \centering
    \begin{tabular}{|c|c|c|c|}
    \hline
    ID & $x_i$ & ID & $x_i$ \\ \hline
     1 &         1 &        8 &         1  \\
     2 &         1 &        9 &         1  \\
     3 &         1 &        10 &         -  \\
     4 &         1 &        11 &         -  \\
     5 &         - &        12 &         0  \\
     6 &         - &        13 &         0  \\
     7 &         0 &        B1 &         -  \\
\hline
    \end{tabular}
    \caption{Fixed and undetermined acceptance indicators of $\Phi^0$ after applying the $MCP$ constraints.}
    \label{table:ex_2_fixed_AIs}
    \end{small}
\end{table}

The fixed acceptance indicators imply 380 units of accepted demand and 240 units of accepted supply, thus Problem 2 must be performed in order to account for this imbalance, i.e. to obtain a surplus of 140 units on the supply side.

However, this is impossible. Without the acceptance of B1, bids 10 and 11 are able to provide only 110 units of supply, and if the B1 is accepted ($Q=150$), the $MCP$ must be at least 50, which implies the full rejection of the demand bids 5 and 6, resulting in over-supply, independent of the acceptance of bids 10 and 11.

In the next subsection we show that these issues may be avoided if other aggregation patterns are used in the examples, and propose the concept of maximally different aggregations in order to obtain different aggregated bid sets, which may help to find the reference solution in the BA-based clearing.


\subsection{Maximally different aggregations}
\label{subsec_MDA}

\begin{figure}[h!]
    \centering
\includegraphics[width=8cm]{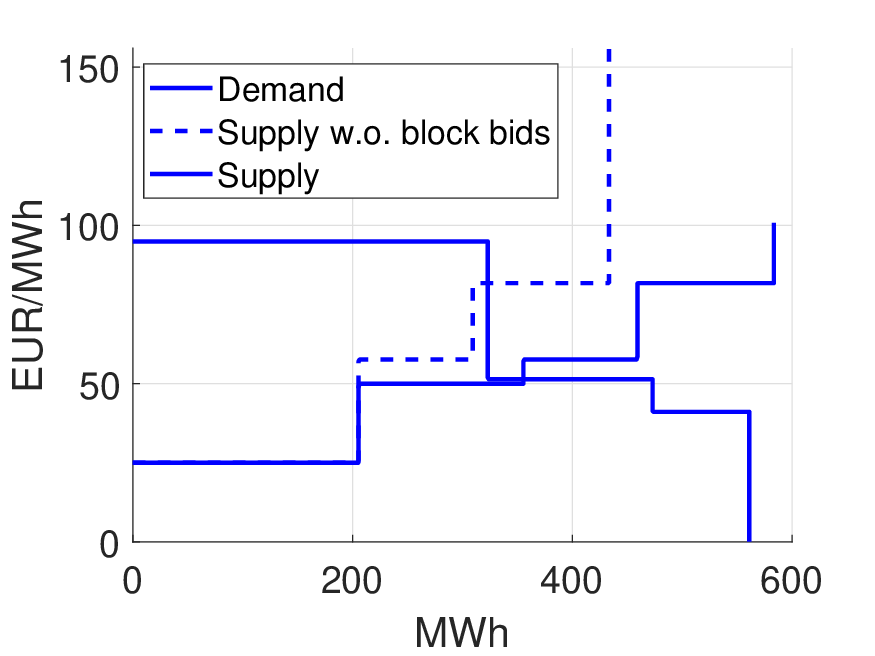}
    \caption{Cumulative supply and demand curves of $\Phi^2$.}
    \label{opt_EX_1}
\end{figure}

Subsections \ref{subsec_example} and \ref{subsec_example_2} demonstrated the potential flaws of the proposed method. However, let us point out some additional properties of the examples. In the case of $\Phi^0$ of example I, instead of the aggregation pattern described in Table \ref{tab_ex_1_subopt_aggr_patt}, let us use the aggregation pattern described in Table \ref{tab_ex_1_aggr_pattern_corrected}, denoted by $\mathcal{A}'$, which is slightly different from the nominal aggregation pattern described in Table \ref{tab_ex_1_subopt_aggr_patt} in step 1, and let $\Phi^2$ stand for the resulting aggregated bid set.
Considering the aggregated bid set $\Phi^2$, step 2 of the BA based clearing will imply the clearing problem depicted in Fig. \ref{opt_EX_1}, resulting in the acceptance of the the block bid B1, and the MCP constraints determined in step 3 will be as $\overline{MCP}=56,~\underline{MCP}=26.6$. Accordingly, problem 2 of step 4 will provide optimal results.

\begin{table}[h!]
\begin{small}
    \centering
    \begin{tabular}{|c|c|c|c|}
    \hline
    ID & $\mathcal{A}'_D(i)$ & ID & $\mathcal{A}'_S(i)$ \\ \hline
     1 &         1 &        8 &         4  \\
     2 &         1 &        9 &         4  \\
     3 &         1 &        10 &        5  \\
     4 &         2 &        11 &        5  \\
     5 &         2 &        12 &        6  \\
     6 &         3 &        13 &        6  \\
     7 &         3 &          &            \\
    \hline
    \end{tabular}
    \caption{Modified aggregation patterns for example I, which results in optimal solution.}
    \label{tab_ex_1_aggr_pattern_corrected}
    \end{small}
\end{table}

Similarly, if in the case of the data set of example II we use the aggregation pattern $\mathcal{A}''$ described in Table \ref{tab_ex_2_aggr_pattern_corrected}, Problem 2 of step 4 will be feasible, and the BA based approach will result in the optimal solution.

\begin{table}[h!]
\begin{small}
    \centering
    \begin{tabular}{|c|c|c|c|}
    \hline
    ID & $\mathcal{A}''_D(i)$ & ID & $\mathcal{A}''_S(i)$ \\ \hline
     1 &         1 &        8 &         4  \\
     2 &         1 &        9 &         4  \\
     3 &         2 &        10 &        5  \\
     4 &         2 &        11 &        5  \\
     5 &         3 &        12 &        6  \\
     6 &         3 &        13 &        6  \\
     7 &         3 &          &            \\
    \hline
    \end{tabular}
    \caption{Modified aggregation pattern for example II, which result in a feasible Problem 2 and optimal solution.}
    \label{tab_ex_2_aggr_pattern_corrected}
    \end{small}
\end{table}

Nevertheless, depending on the aggregation pattern, the BA based two-stage clearing method may provide suboptimal or no results (in the case of an infeasible Problem 2).
In this paper we propose a computational approach to overcome this issue: We define other aggregations as well in addition to the nominal aggregation $\mathcal{A}^N$. The idea behind our approach is that if the breakpoints of the cumulative supply and demand curves resulting from the aggregated bids are different, the estimations regarding the $MCP$ intervals of periods will be different (see subsection \ref{subsec_step_3} for details). Computations based on these different aggregations may run parallel with the nominal aggregation based method. When the computations of each strands (or threads) are done, we evaluate their outcomes regarding final feasibility and the resulting TSW.

\subsubsection{Definition of maximally different aggregations}

The maximally different aggregations are derived period-wise for the supply and demand side. 
Our approach for constructing additional aggregations will be to derive $R$ random, potentially different aggregations based on $\mathcal{A}^N$, and use the ones among them which are the 'most different' from the nominal aggregation, based on a max-min type measure regarding the breakpoints of the cumulative supply or/and demand curves resulting from the aggregated bids.
Let us note that in general this method does not ensure that the max/min distance of the breakpoints of cumulative curves will be maximal from the original case regarding all possible aggregations, with the same number of aggregated bids. Calculating such an aggregation would require intensive combinatorial optimization in the case of large bid numbers, and as we will see this simple approach will be sufficient for our aims.

We define the algorithm for the demand side (for the supply side is straightforward):
\begin{enumerate}
    \item Consider the sets of original standard demand bids ($SD^0$) and the vectors $\mathcal{A}^N_{D,t}$ for each period $t$.
    \item We generate $R$ subdivisons of the original bids in $SD_t$.
    A single subdivision $r \in \{1,\ldots,R\}$, and the implied aggregation $\mathcal{A}^r$ is generated as follows. Let us define the following notations. $n_{D,t}=|SD_t|$ denotes the number bids in $SD^0$ for period $t$ and let $N_{D,t}=\max \left( \mathcal{A}^N_{D,t} \right)$ be the number of aggregated demand bids for period $t$. We pick $N_{D,t}-1$ random integers in the range of $1$ to $n_{D,t}$, denoted by $(b_0=)~1 < b_1 < b_2 < \ldots < b_{N_{D,t}-1} < n_{D,t}~ (=b_{N_{D,t}})$. 
    These chosen numbers represent the 'bin limits' of each aggregated bid (bids with indices between $b_i$ and $b_{i+1}$ will be joined into a single aggregated bid, as described by eq.  (\ref{eq_random_aggr}) \footnote{In addition, we have to consider that if there are bids with equal price, they will be joined in the same aggregated bid}.
    \begin{align}
    &\mathcal{A}^r_{D,t}(1)= \ldots = \mathcal{A}^r_{D,t}(b_1)=1\nonumber \\
    &\mathcal{A}^r_{D,t}(b_1 + 1)= \ldots = \mathcal{A}^r_{D,t}(b_2)=2\nonumber \\
    & \vdots \nonumber \\
    &\mathcal{A}^r_{D,t}(b_{N_{D,t}-1}+1)= \ldots= \mathcal{A}^r_{D,t}(n_{D,t})=N_{D,t} \label{eq_random_aggr}
    \end{align}
    
    After calculating the implied aggregated bids according to eq. (\ref{eq:aggregation}), we will have $R$ different aggregation patterns (or aggregations) for the original bid set (denoted by $\mathcal{A}_{D,t}^1, \ldots, \mathcal{A}_{D,t}^R$ respectively).

    \item Calculate the cumulative demand quantities ($Q_{C,D,t}^r$) for each aggregation $r$ (for each period $t$). These values will be equivalent to the \emph{breakpoints} of the cumulative demand curves resulting from the aggregated bids.
    \begin{equation}
        Q_{C,D,t}^r(j) = \sum_{m = 1}^{j} Q^r_{m}~~~m \in ASD_t,
        \label{eq_cumulative_quantities}
    \end{equation}
    where  $Q^r_{m}$ is the quantity of the aggregated (demand) bid $m$ resulting from the aggregation vector $\mathcal{A}^r_{D,t}$.
    \item 
    We can define the \textit{distance} ($\Delta$) between two aggregations ($\mathcal{A}^r_{D,t}$ and $\mathcal{A}^p_{D,t}$): as the absolute value of the difference of their two closest breakpoints. Formally, 
    \begin{equation}
    \Delta \left( \mathcal{A}^r_{D,t}, \mathcal{A}^p_{D,t} \right) = \min\limits_{k, l} \left\{  |Q^r_{C,D,t}(k) - Q_{C,D,t}^p(l) | \right\}~~.
    \end{equation}
    \item Finally we choose the aggregation whose distance from the nominal aggregation is the maximal. We call this aggregation \emph{maximally different aggregation, with respect to demand for period $t$}, and denote it by $\mathcal{A}^{MD}_{D,t}$. 
    \begin{equation}
    \mathcal{A}^{MD}_{D,t} = \left\{ \mathcal{A}^r : \max\limits_{r} \left\{ \Delta \left( \mathcal{A}^r, \mathcal{A}^N \right) \right\}, ~  r \in \{1, ..., R\} \right\}
\end{equation}
    Let us note that finding the most different possible aggregation, according to the proposed or to other norm may be approached by different methods as well.

\end{enumerate}

Based on the determined maximally different aggregations for the supply and demand side, we define 3 additional aggregations, as described by eq. (\ref{eq_MD_aggregations}).
In $\mathcal{A}^1$, the demand side is maximally different compared to $\mathcal{A}^N$, while the supply side bids are aggregated according to $\mathcal{A}^N$. $\mathcal{A}^2$ is the dual of $\mathcal{A}^1$ in the sense that for supply side bids the maximally different aggregation pattern is used and for demand side we use $\mathcal{A}^N$. In the case of $\mathcal{A}^3$, both sides are aggregated based on the maximally different aggregations.

\begin{align}
&\mathcal{A}^1_{D,t}=\mathcal{A}^{MD}_{D,t},~~A^1_{S,t}=\mathcal{A}^{N}_{S,t}~~\forall t \nonumber \\
&\mathcal{A}^2_{D,t}=\mathcal{A}^{N}_{D,t},~~\mathcal{A}^2_{S,t}=\mathcal{A}^{MD}_{S,t}~~\forall t\nonumber \\
&\mathcal{A}^3_{D,t}=\mathcal{A}^{MD}_{D,t},~~\mathcal{A}^3_{S,t}=\mathcal{A}^{MD}_{D,t}~~\forall t \label{eq_MD_aggregations}
\end{align}

\begin{figure}[h!]
        \centering
        \begin{subfigure}[b]{0.47\textwidth}
    \includegraphics[width=\textwidth]{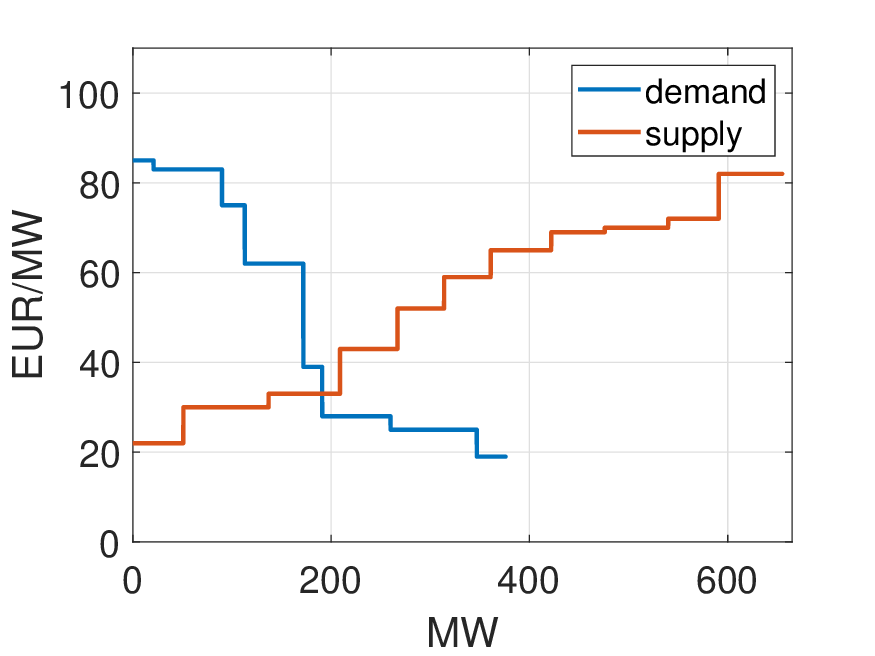}
\caption{Cumulative supply and demand curves of original bids. \label{fig:aggr_orig}}
        \end{subfigure}
        \vskip\baselineskip
        \begin{subfigure}[b]{0.47\textwidth}
            \centering
            \includegraphics[width=\textwidth]{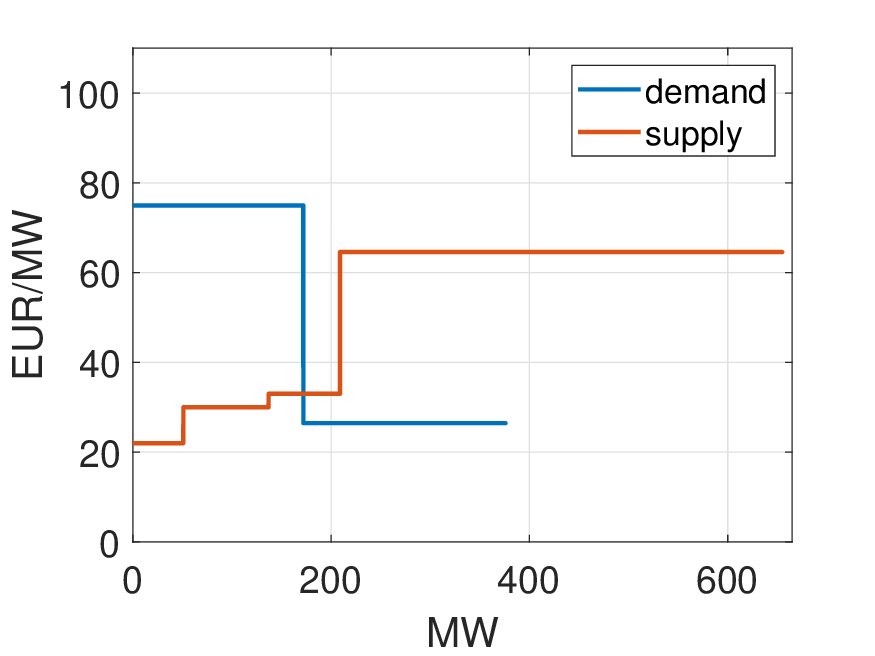}
            \caption[]{$\mathcal{A}^N$: Nominal aggregation.}    
            \label{fig_A0}
        \end{subfigure}
        \begin{subfigure}[b]{0.46\textwidth}  
            \centering 
            \includegraphics[width=\textwidth]{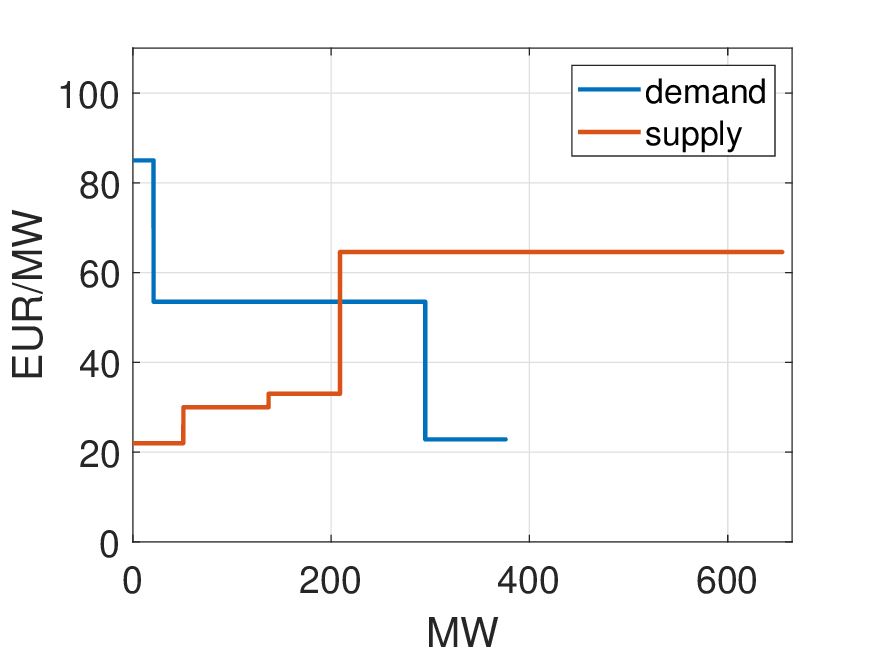}
            \caption[]{$\mathcal{A}^1$: Aggregation with maximally different demand side.}    
            \label{fig_A1}
        \end{subfigure}
        \vskip\baselineskip
        \begin{subfigure}[b]{0.46\textwidth}   
            \centering 
            \includegraphics[width=\textwidth]{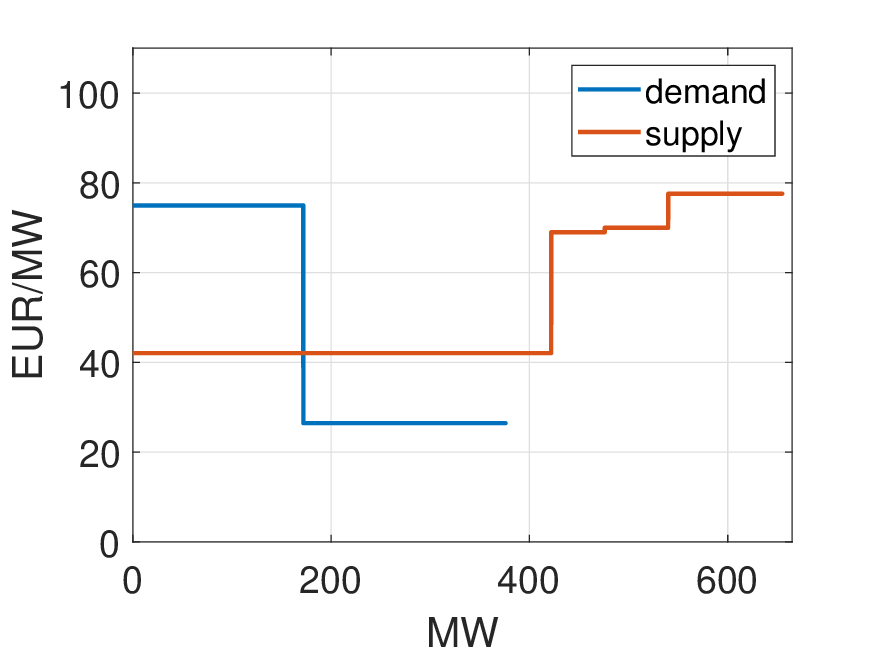}
            \caption[]{$\mathcal{A}^2$: Aggregation with maximally different supply side.}  
            \label{fig_A2}
        \end{subfigure}
        \hfill
        \begin{subfigure}[b]{0.46\textwidth}   
            \centering 
            \includegraphics[width=\textwidth]{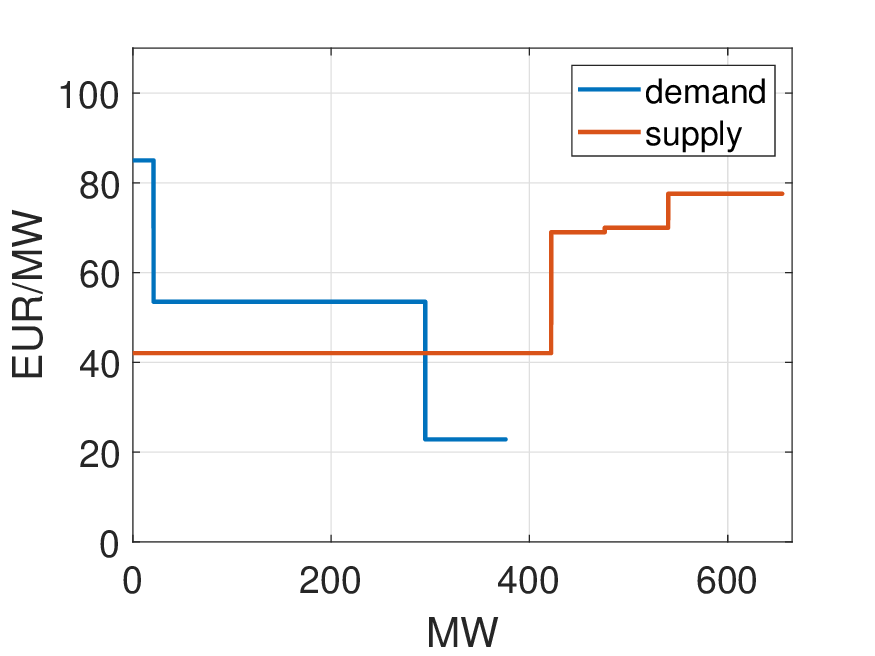}
            \caption[]{$\mathcal{A}^3$: Aggregation where both sides are maximally different compared to $\mathcal{A}^N$.}    
            \label{fig_A3}
        \end{subfigure}
\end{figure}

 Figs \ref{fig_A0} -- \ref{fig_A3} demonstrate the 4 different aggregated curves in the case of a simple 1-period example, with the original bid set depicted in Fig \ref{fig:aggr_orig}. 


\section{Numerical results and discussion}
\label{sec_results}

\subsection{Data and simulations}
\label{subsec_data_and_simulations}

Bid sets for numerical tests have been generated based on the data set of \cite{biskas_dataset}. While the data described in \cite{biskas_dataset} assumes 14 bidding zones with interconnections, in the current setup these have been merged into a single zone. 
9 different setups have been generated from the original data, where the number of time periods, and the number of block bids vary.

Since in the original data the prices of block orders were usually significantly below the resulting MCPs, paradox rejection and thus the potential problems related to the proposed approach were not prevalent. In order to make the data set more challenging for the proposed approach, the price values of block orders have been increased by 25\% to make paradox rejection and related phenomena more common.
In addition, in cases where the total quantity of supply was 2-3 times more compared to the total quantity of demand, a random subset of supply bids has been removed from the bid set to make the total resulting supply and demand quantities more symmetric, and the
number of standard supply and demand bids have been curtailed to 140 in all test cases per period.

Based on the reference data described above and on method described in \cite{csercsik2022synthesis}, 100 different bid sets have been generated for testing purposes for each setup.


\subsection{Test setups}
\label{subsec_setups}

As it had been already mentioned, 9 different setups have been used during the numerical tests. Table \ref{tab:setups} summarizes the parameters of setups.

\begin{table}[h!]
    \centering
    \begin{tabular}{|c|c|c|c|}
    \hline
         \thead{Setup ID}& \thead{n.o. time \\ periods} & \thead{n.o. standard \\bids per period } & \thead{n.o. block \\bids} \\ \hline
         1 & 12 &  280  & 262 \\
         2 & 12 &  280  & 524 \\
         3 & 12 &  280  & 1048 \\
         4 & 18 &  280  & 262 \\
         5 & 18 &  280  & 524 \\
         6 & 18 &  280  & 1048 \\
         7 & 24 &  280  & 262 \\
         8 & 24 &  280  & 524 \\
         9 & 24 &  280  & 1048 \\   
         \hline
    \end{tabular}
    \caption{Market setups used in computational tests}
    \label{tab:setups}
\end{table}

\subsection{Reliability}

As we have already discussed, the BA based clearing method may encounter suboptimality or infeasibility-related problems, which originate from the phenomena of paradox rejection of block bids. First we examined the success rate of the proposed method, in the case of the various setups.
The outcome of the BA based method has been compared to the solution of the original clearing problem (considering the clearing of the original bid set $\Phi^0$ without aggregation). A particular case was taken into account as success, if the stage 2 problem of the proposed method has been proven to be feasible, and the resulting TSW was no less than 99.999 \% of the original value (excluding infeasibility and suboptimality). Table 
\ref{tab:succes_rate} summarizes the results.

\begin{table}[h!]
    \centering
    \begin{tabular}{|c|c|c|} \hline
    \thead{setup ID}&\thead{success\\rate} & \thead{feasibility \\ rate} \\ \hline
    1 &         1 (0.98) &         1 (0.99)  \\
    2 &         1 (0.99) &         1 (1)  \\
    3 &         1 (0.92) &         1 (1)  \\
    4 &         1 (0.95) &         1 (1)  \\
    5 &         1 (0.97) &         1 (1)  \\
    6 &         1 (0.98) &         1 (1)  \\
    7 &         1 (0.99) &         1 (1)  \\
    8 &      0.99 (0.97) &         1 (1)  \\
    9 &         1 (0.96) &         1 (1)  \\\hline
    \end{tabular}
    \caption{Success rate and feasibility rate of the BA based method using 4 different aggregations in parallel. The numbers in parentheses are the success rates and feasibility rates, if only the nominal aggregation is used (not the 4 different aggregations in parallel).}
    \label{tab:succes_rate}
\end{table}

As Table \ref{tab:succes_rate} clearly shows, using 4 different aggregations in parallel significantly increases the success rate of the proposed method (if the parallel approach is used, a test case is considered successful, if any of the threads is successful).


It can be see in Table \ref{tab:succes_rate}, that while the infeasibility of the second clearing step in step 4 of the algorithm is theoretically possible as demonstrated in Example II in subsection \ref{subsec_example_2}, it is not prevalent in the case of realistic scenarios with high number of standard bids and block orders. 

\subsection{Computational gain}

\begin{figure}[h!]
    \centering
    \includegraphics[scale=0.7]{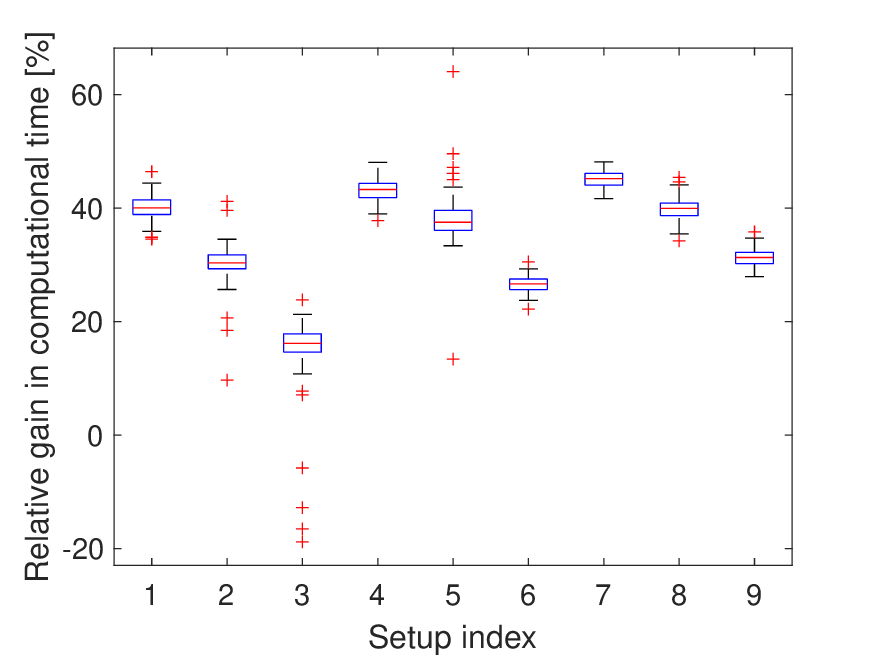}
    \caption{Distribution of the values of computational gain}
    \label{fig:comp_gain}
\end{figure}

Table \ref{tab:comp_gain} summarizes the values corresponding to the relative reduction in computational time in the case of the various test setups. 
The computational time in the case of the parallel BA-based method was interpreted as the time needed until the slowest thread also finished. Figure \ref{fig:comp_gain} depicts the distribution of these values. In the box plots, the central mark is the median, while the edges of the box are the 25th and 75th percentiles respectively. The whiskers extend to the most extreme data points which are considered not to be outliers, and the outliers are plotted individually with red crosses. As it can be seen in the case of setup 3, where the ratio of block orders is the highest compared to the total number of standard bids, in extreme cases the BA-based method may be slower as well compared to the original clearing, but this phenomena is very rare and does not occur, if the ratio of standard bids vs. block orders is high enough.

\begin{table}[h!]
    \centering
    \begin{tabular}{|c|c|}
    \hline
    Setup ID & comp. gain [\%] \\ \hline
    1 &      40.2  \\
    2 &      30.2  \\
    3 &      15.1  \\
    4 &      43.2  \\
    5 &        38  \\
    6 &      26.5  \\
    7 &        45  \\
    8 &      39.8  \\
    9 &      31.3  \\
    \hline
    \end{tabular}
    \caption{Average computational gain of the BA based method.}
    \label{tab:comp_gain}
\end{table}

The statistics corresponding to the absolute values of the computational times are described in Table \ref{tab:t_abs_values} in Appendix B.

\subsection{Implementation}
The numerical tests have been performed on a desktop computer, with an Intel core i5 processor @2.9 GHz and 16GB of RAM. The optimization problem of the market clearing has been solved using the GUROBI solver in AMPL environment. The BA-based clearing process has been implemented in MATLAB, from where the AMPL has been called.

\subsection{Discussion}
\label{subsec_discussion}

\subsubsection{Computational gain and problem size}

The computational gain is implied by the reduction in the effective problem sizes.
For example, in the case of setup 1, the number of standard bids is 3360 (=280*12). Using the nominal aggregation with standard parameters in the clustering method results in appr.
1475 -- 1492 standard bids in the clearing of step 2, while after the application of the derived MCP constraints in the presolve step of Problem 2 in step 4, the number of undetermined acceptance indicator variables of standard bids ranges in average from 58 to 102. The rate of reduction is very similar in the case of other test setups as well (the number of standard bids is reduced to appr. 45\%  of the original value in step 2 due to the aggregation, and the number of undetermined acceptance variables corresponding to standard bids is about 2-3\% of the original value in step 4 due to the application of the MCP constraints).

Two trends are observable in the results summarized in Table \ref{tab:comp_gain}. On the one hand, if the number of standard bids is increased, ceteris paribus (consider e.g. the scenarios 2, 5 and 8), the computational gain increases as well. On the other hand, as the number of block orders is increased, the computational gain becomes smaller. The explanation for the latter trend is that block bids are not subject to aggregation, and the integer variables implied by them strongly influence the execution time of the clearing algorithm.

Let us note that there is a minimum bid number, below which the method does not bring any computational gains in average.
In the case e.g. of a reference setup with 5 periods, 50 standard bids per period and 262 block bids, 
the execution time of the full-scale problem is usually already below the computational time of the BA based method, however the exact values may of course vary, depending on the individual bid set.

\subsubsection{Potential application of the proposed framework}

As we have seen, the proposed method is not always able to provide a solution, and even if a feasible solution is obtained it may be suboptimal -- even if with a relatively small optimality gap.
Although (if computing capacity allows) increasing the number of aggregation patterns, which run in parallel, over four may make these phenomena even more rare, it is likely that these flaws cannot be completely eliminated by such techniques.
However, if the problem size is large enough, the BA based method provides an optimal solution in the majority of the cases significantly (by 20 -40 \%) faster compared to the 'full scale' clearing. Taking this into account, in time-critical applications, where the full scale clearing may fail to provide results within the allowed time frame, it may be advisable to use the original clearing method together with the BA based algorithm in a parallel fashion with the same input data.
As Tables \ref{tab:succes_rate} and \ref{tab:comp_gain} show, the BA based method is able to provide (at least a suboptimal) solution relatively fast in the vast majority of cases, which may be valuable if the full scale solver fails to do it so within the allowed time frame.

\section{Conclusions and future work}
\label{sec_conclusions}

Although there are recent literature results proposing the Benders-decomposition approach for the solution of the day-ahead market clearing problem \cite{dourbois2014european,madani2015computationally}, which also allows the utilization of parallel computing resources at some level, the method of BA-based clearing proposed in this article provides an opportunity to utilize parallel computing architectures for the day-ahead market clearing problem in a more simple and intuitive way. Furthermore, in the proposed setup, parallelization is carried out in a controlled manner, and since the number of possible different aggregations is very high (the four different aggregation approaches may be also combined on the level of periods), practically any capacity of parallel computation architecture may be efficiently utilized, if available. In addition, as already mentioned, the individual clearing steps of the proposed algorithm are standard market clearing problems, thus they may also be carried out by using any other already tested efficient techniques as well (e.g. by Benders-decomposition), combining the approaches to reach an even higher computational performance.

 \subsection{Future work}

The model proposed in this paper assumed only block orders as non-convex orders and only a single market zone. 
While the generalization of the approach for further complex bid types, like minimum income condition orders \cite{divenyi2021investigating} is quite straightforward, as they may be considered in a way similar to block orders, i.e. not to include them in the aggregation, but to account for their possible rejection according to the derived MCP-estimates, the possible application of the proposed method in a market coupling setting of multiple zones is less trivial. One may of course aggregate standard bids period- and zone-wise, but it is possible that the interaction of the estimated zonal MCP-ranges and line loads will imply further complex problems in the case of the BA based clearing, thus such an application setting should be in the focus of future studies.

\section{Acknowledgements}
The authors thank P\'{e}ter M\'{a}rk S\H{o}r\'{e}s for his advices to improve the manuscript.
This work has been supported by the Fund FK 137608 of the Hungarian National Research, Development and Innovation Office. 

\section*{Bibliography}
\bibliographystyle{plain}

\bibliography{bibek.bib}

\section*{Appendix A}
\label{Appendix A}
The clustering algorithm used for the determination of the nominal aggregation pattern is as follows.
\begin{enumerate}
    \item Calculate the distance (difference) between every pair of bid prices (demand and supply bide separately):  in case of $n$ bids we will have $n \frac{n-1}{2}$ distance values.
    \item Forming binary clusters of the bids based on their distances (linking bids that are close to each other in price)
    \item Link these newly formed clusters to each other to create bigger clusters until all the bids in the original data set are linked together in a hierarchical tree.
    In this step the algorithm stores the (cophonetic) distances of all the clusters on the same hierarchical level from each other.
    \item Determine the cluster divisions \newline
    One way to complete this is to compare the height of each link in a cluster tree with the heights of neighboring links below it in the tree. A link that is approximately the same height as the links below it indicates that there are no distinct divisions between the objects joined at this level of the hierarchy. These links are said to exhibit a high level of consistency, because the distance between the objects being joined is approximately the same as the distances between the objects they contain.
    On the other hand, a link whose height differs noticeably from the height of the links below it indicates that the objects joined at this level in the cluster tree are much farther apart from each other than their components were when they were joined. This link is said to be inconsistent with the links below it.
    We can set a cutoff parameter for the clustering above which the algorithm does not group the given objects together.
    In our case this parameter is set to the median of the inconsistency coefficients to reach enough, but not too many clusters of bids.
\end{enumerate}

\section*{Appendix B}
The absolute values of the numerical results of the simulation time values are as follows.

\begin{table}[h!]
    \centering
    \begin{tabular}{|c|c|c|c|c|c|c|}
    \hline
    Setup ID & $\overline{t}_{ref}$ & $\overline{t}_{BA}$ & med(${t}_{ref}$) & med(${t}_{BA}$) & std(${t}_{ref}$) & std(${t}_{BA}$)  \\ \hline
    1 & 44.3 &      26.5 &      44.3 &      26.5 &     0.354 &      0.92  \\ \hline
    2 &47.6 &      33.2 &      47.9 &      33.3 &      1.52 &      1.23  \\ \hline
    3 &53.3 &      45.1 &      53.8 &        45 &      2.13 &      2.66  \\ \hline
    4 & 65.7 &      37.3 &      65.8 &      37.2 &     0.499 &      1.32  \\ \hline
    5 & 65.7 &      40.7 &      69.5 &      42.7 &      6.73 &      5.03  \\ \hline
    6 & 77.3 &      56.8 &      77.3 &      56.7 &     0.193 &      1.14  \\ \hline
    7 & 88.2 &      48.5 &      88.2 &      48.3 &     0.237 &      1.25  \\ \hline
    8 & 91.9 &      55.3 &      92.4 &      55.3 &      1.44 &      1.69  \\ \hline
    9 & 101 &      69.4 &       101 &      69.4 &      0.31 &      1.55  \\ \hline
    \end{tabular}
    \caption{Average, median and standard deviation values of the reference computational time [s] ($t_{ref}$ - with no aggregation) and the computational time in the case of the bid-aggregation based method ($t_{BA}$)}
    \label{tab:t_abs_values}
\end{table}






\end{document}